%% file: latpulsars.tex
\documentclass[graybox]{svmult}

\usepackage{url}
\usepackage{hyperref}
\usepackage{natbib}
\usepackage{mathptmx}       
\usepackage{helvet}         
\usepackage{courier}        
\usepackage{type1cm}        
\usepackage{makeidx}         
\usepackage{graphicx}        
\usepackage{multicol}        
\usepackage[bottom]{footmisc}

\makeindex             

\begin{document}

\title*{Pulsar Results with the {\it Fermi} Large Area Telescope}
\author{Paul S. Ray and Pablo M. Saz Parkinson for the {\it Fermi} LAT Collaboration, \\ the 
LAT Pulsar Timing Consortium, and the LAT Pulsar Search Consortium}
\institute{Paul S. Ray \at Naval Research Laboratory, 4555 Overlook Ave., SW, Washington 
DC 20375-5352 USA \\ \email{paul.ray@nrl.navy.mil}
\and Pablo M. Saz Parkinson \at Santa Cruz Institute for Particle Physics, University of 
California, Santa Cruz, CA 95064 USA \\ \email{pablo@scipp.ucsc.edu}}
\authorrunning{P. S. Ray and P. M. Saz Parkinson}
\maketitle

\abstract{The launch of the \textit{Fermi Gamma-ray Space Telescope} has heralded a new 
era in the study of gamma-ray pulsars. The population of confirmed gamma-ray pulsars has 
gone from 6--7 to more than 60, and the superb sensitivity of the Large Area 
Telescope (LAT) on \textit{Fermi} has allowed the detailed study of their spectra and light 
curves. Twenty-four of these pulsars were discovered in blind searches of the gamma-ray 
data, and twenty-one of these are, at present, radio quiet, despite deep radio follow-up 
observations. In addition, millisecond pulsars have been confirmed as a class of 
gamma-ray emitters, both individually and collectively in globular clusters. Recently, 
radio searches in the direction of LAT sources with no likely counterparts have been highly 
productive, leading to the discovery of a large number of new millisecond pulsars. Taken together, these 
discoveries promise a great improvement in the understanding of the gamma-ray emission 
properties and Galactic population of pulsars. We summarize some of the results stemming 
from these newly-detected pulsars and their timing and multi-wavelength follow-up observations.}

\abstract*{The launch of the \textit{Fermi Gamma-ray Space Telescope} has heralded a new 
era in the study of gamma-ray pulsars. The population of confirmed gamma-ray pulsars has 
gone from 6--7 to more than 60, and the superb sensitivity of the Large Area 
Telescope (LAT) on \textit{Fermi} has allowed the detailed study of their spectra and light 
curves. Twenty-four of these pulsars were discovered in blind searches of the gamma-ray 
data, and twenty-one of these are, at present, radio quiet, despite deep radio follow-up 
observations. In addition, millisecond pulsars have been confirmed as a class of 
gamma-ray emitters, both individually and collectively in globular clusters. Recently, 
radio searches in the direction of LAT sources with no likely counterparts have been highly 
productive, leading to the discovery of a large number of new millisecond pulsars. Taken together, these 
discoveries promise a great improvement in the understanding of the gamma-ray emission 
properties and Galactic population of pulsars. We summarize some of the results stemming 
from these newly-detected pulsars and their timing and multi-wavelength follow-up observations.}

\section{Introduction}
\label{sec:intro}

\subsection{Gamma-ray Pulsars in the Year 2000}
\label{subsec:EGRET}

Ten years ago, on 4 June 2000, the \textit{Compton Gamma Ray Observatory} (CGRO) was de-orbited, ending 
nine years of operation, during which it revolutionized gamma-ray astronomy. In 
particular, the Energetic Gamma Ray Experiment Telescope (EGRET) surveyed the sky at 
energies $>100$ MeV with much better sensitivity than previous experiments. The landmark 
Third EGRET (3EG) Catalog \citep{3rdCat} reported the characteristics of 271 
gamma-ray sources. The largest class of identified sources were blazars, with 66 sources, 
followed by 5 pulsars (Crab, Vela, Geminga, PSR B1055$-$52, and PSR B1706$-$44), 1 solar 
flare, the Large Magellenic Cloud, and one probable radio galaxy (Centaurus A).   
Interestingly, the majority of the 3EG sources (170 of them) were not associated with 
any known classes of gamma-ray emitting objects. It was widely believed that a large 
number of the unidentified EGRET sources, particularly along the Galactic plane, could be 
pulsars \citep[e.g.][]{yad95}, and several radio pulsars were, in fact, discovered by searching the 
error circles of EGRET unidentified sources \citep[e.g.][]{Halpern2001a,Roberts2002}. Further work 
on EGRET data revealed one more high-confidence pulsar (PSR B1951+32) and several candidates, including 
one millisecond pulsar \citep{Kuiper2000}. An excellent observational summary of what was known about gamma-ray 
pulsars at the end of the EGRET era was presented by \citet{thom01}. It is also worth noting 
that a 7th gamma-ray pulsar (PSR B1509$-$58) was detected by the COMPTEL experiment up to 10 MeV 
\citep{Kuiper1999}, though it was never seen with EGRET. 
Pre-launch predictions of the number of gamma-ray pulsars that {\it Fermi} LAT would detect (as well as the 
fraction of those that would be radio quiet) are highly dependent on the assumed gamma-ray emission model, 
ranging from a few tens to many hundreds~\citep[e.g.][]{Ransom07}, with the larger number (and fraction of radio-quiet pulsars) 
usually predicted by outer-magnetosphere models, where the gamma-ray beam is expected to be 
broader~\citep{Jiang06,Harding07}. It should be noted that the detection of a gamma-ray pulsar, in 
this context, does not necessarily imply the detection of its pulsations; most models, for example, 
``predict" that EGRET detected far more than the 6 gamma-ray pulsars for which high-confidence pulsations 
were actually observed, a view that is supported by the subsequent detection of pulsations from many formerly 
unidentified EGRET sources by the LAT. 

\subsection{{\it Fermi} and {\it AGILE}}

After almost a decade without an orbiting GeV telescope, two new satellites were 
launched in 2007--2008, ushering in a new era of gamma-ray astronomy. \textit{AGILE} (an 
Italian acronym for \textit{Astro-rivelatore Gamma a Immagini LEggero}) was launched on 
23 April 2007 and the \textit{Fermi Gamma-ray Space Telescope} (formerly GLAST) was 
launched on 11 June 2008. The prime instruments on both spacecraft are pair production 
gamma-ray telsecopes, like EGRET. However, instead of a gas spark chamber, they employ 
more modern solid-state silicon strip detectors to track the gamma-ray and particle events. 
While \textit{AGILE} had a 14-month head start on \textit{Fermi}, and has made many important 
contributions, it represents a modest improvement in sensitivity compared to EGRET. In 
this paper we focus on the pulsar results made possible by the enormous leap in 
sensitivity afforded by \textit{Fermi}.

The primary instrument on \textit{Fermi} is the Large Area Telescope (LAT) \citep
{LATinstrument}.  The LAT is a pair conversion gamma-ray telescope where incoming 
gamma rays are converted to electron-positron pairs in a set of tungsten foils. The 
resulting electron-positron pair and shower of secondary particles are tracked by a stack of 
silicon strip detectors to determine the incident direction of the photon before the 
energy of the shower is recorded in a CsI calorimeter. The instrument is wrapped in a 
segmented anti-coincidence detector that aids in the separation of events due to charged 
particles from those resulting from photons. This is critical because charged particle 
events outnumber photon events by a factor of 10$^4$. The LAT is sensitive to photons in 
the energy range 30 MeV to $>300$ GeV, with an effective area of $\sim8000$ cm$^2$ at 1 
GeV. The point spread function is $\sim 0.8^\circ$ at 1 GeV and is a strong function of 
energy, scaling like $E^{-0.8}$ until the resolution becomes limited by position 
resolution in the tracker at about 0.07$^\circ$\footnote{This is the individual photon angular resolution. Bright sources can be localized more precisely via centroiding.}. Compared to EGRET, the LAT represents a 
major improvement in effective area, field of view, and angular resolution. In addition, 
it operates in a sky survey mode which avoids loss of observing efficiency from Earth 
occultations and covers the sky nearly uniformly every two orbits ($\sim 3$ hours). 
These characteristics give the LAT unprecedented sensitivity for discovery and study of 
gamma-ray pulsars. The First {\it Fermi} LAT catalog \citep[1FGL;][]{1FGL} of 1451 gamma-ray sources 
detected during the first 11 months of science operations contains 56 sources that have 
been firmly identified as pulsars through their gamma-ray pulsations. Several additional gamma-ray pulsars 
have been identified since the release of the catalog, bringing the total number to more than 60. In 
the following sections, we describe the various populations of gamma-ray pulsars being explored by the LAT, and 
the different techniques employed in their detection, as well as some of the new insights being gained
through these new findings. We end with a brief summary and some thoughts on the future goals and
expectations for pulsar astrophysics with the LAT in the coming years.  

\section{The EGRET Pulsars in Exquisite Detail}

The EGRET experiment represented a major improvement relative to previous gamma-ray missions 
(e.g. SAS-2 and COS-B). In addition to increasing the number of high-confidence gamma-ray 
pulsars from 2 to 6, the higher sensitivity of EGRET led to a better understanding of the 
known gamma-ray pulsars (at the time, only the Crab and Vela). Similarly, the LAT, with its improved 
sensitivity and broader energy range is not only enabling the discovery of a large number of new 
gamma-ray pulsars, but is also greatly expanding our knowledge of the previously known EGRET-detected 
pulsars. Because these pulsars are among the brightest known gamma-ray sources, 
the LAT is able to accumulate enough statistics to allow for detailed (and phase-resolved) 
spectral analyses, in many cases answering some questions left over from the EGRET era, or 
challenging some of the previous EGRET results which in most cases were based on limited statistics.

\begin{figure}[b]
\sidecaption
\includegraphics[width=75mm]{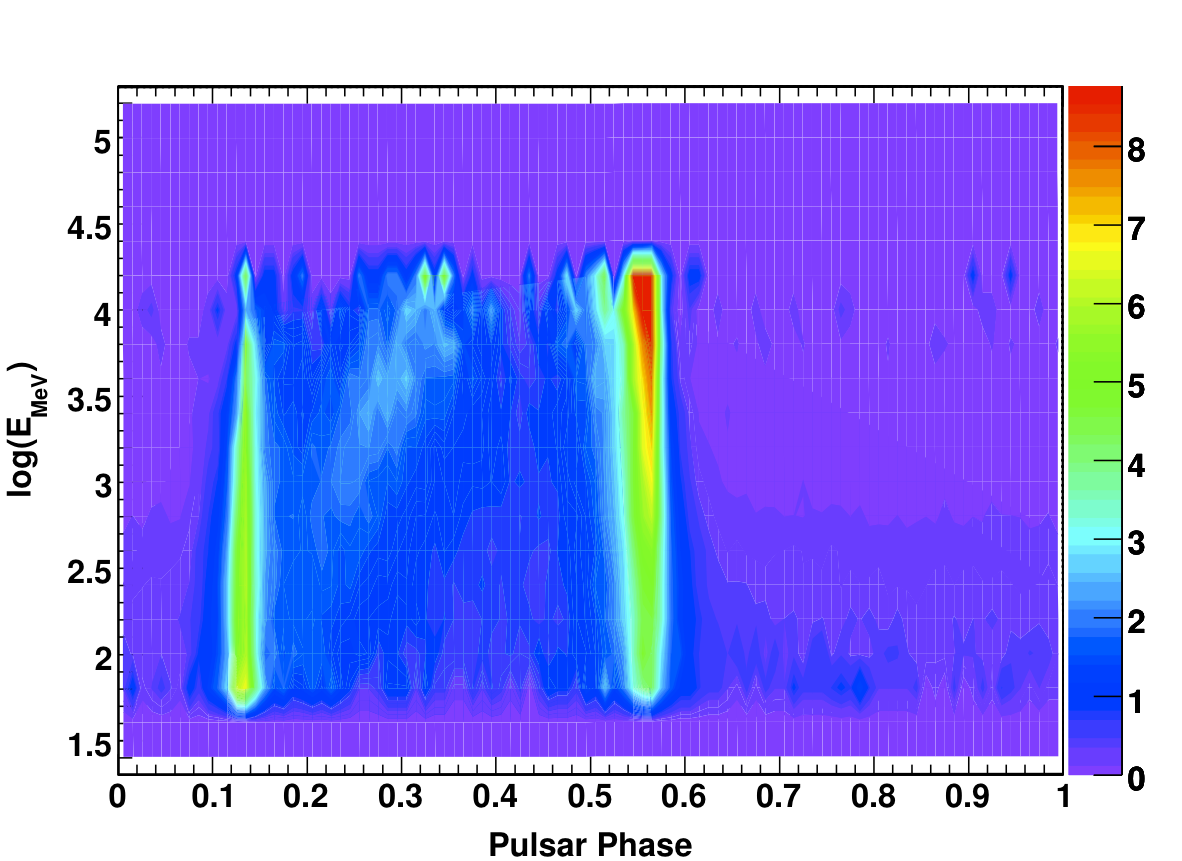}
\caption{Pulse profile of the Vela pulsar, as a function of energy. The different 
behavior of the two main peaks is evident. A third peak is seen to appear at higher 
energies, with its position shifting in phase, as a function of energy \citep[from][reproduced by permission of the AAS]{LATVELA2}.}
\label{fig:vela_third_peak}       
\end{figure}

Early LAT observations of Vela, the brightest steady gamma-ray source, confirmed some of the
basic features of this pulsar: It has two asymmetric peaks that evolve differently with 
energy, and a phase-averaged spectrum well modeled by a hard power-law with an exponential 
cutoff in the 2--4 GeV energy range. In addition, the much better statistics and time resolution of 
the LAT data reveal pulse structures as fine as 0.3 ms, and a hitherto unknown third peak in the 
light curve, which evolves with energy (see Figure~\ref{fig:vela_third_peak}). Spectral fits 
to the LAT data suggest that a simple exponential cutoff is preferred over a super-exponential one, 
indicating that outer-magnetosphere emission models are favored over polar cap type 
models~\citep{LATVELA}. More recent results on Vela, using 11 months of observations, show detailed 
phase-resolved features which confirm the EGRET results on the spectral evolution of the two main peaks. 
In addition, while the first peak is seen to fade at higher energies, the newly-discovered third peak, along 
with the second peak, are present up to the highest detected pulsed energies~\citep{LATVELA2}. 

\begin{figure}
\sidecaption[t]
\includegraphics[width=70mm]{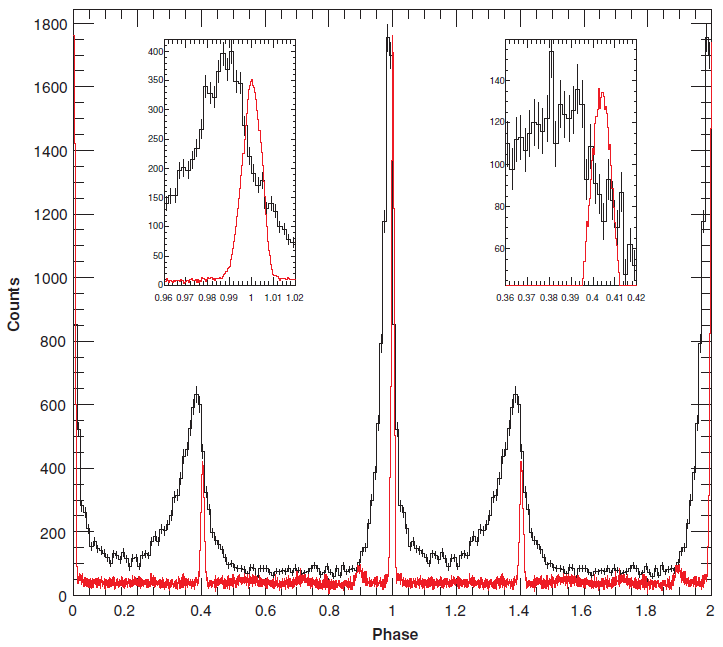}
\caption{Folded light curve of the Crab pulsar \citep[from][reproduced by permission of the AAS]{LATCRAB}. The statistics 
provided by the LAT allow us to observe structure in the light curve with incredible 
precision. For example, it is now clear that the main radio pulse (red) and the gamma-ray 
pulse (black) do not line up, and in fact are separated by approximately 0.3 ms.}
\label{fig:crab}       
\end{figure}

LAT results on the Crab pulsar confirm that it shares many of the properties of Vela, with two asymmetric
peaks evolving differently with energy. The Crab pulsar spectrum is also best modeled with a power law 
with an exponential cutoff, but the cut-off energy in this case is much higher than Vela ($\sim$6 GeV), with  
pulsed gamma-ray photons being detected at least up to $\sim$20 GeV~\citep{LATCRAB}. One of the new features 
uncovered by the LAT is an apparent phase shift between the main radio peak and the first gamma-ray peak. 
Previously, it was thought that these two were aligned, but the fine time resolution of the LAT allows us to 
determine that the first gamma-ray peak leads the main radio pulse by ($281\pm12\pm21$) $\mu$s 
(see Figure~\ref{fig:crab}).

In addition to being the second brightest non-variable source in the GeV sky, Geminga was the first known radio-quiet 
gamma-ray pulsar. As such, it cannot be timed in radio and until now, a good timing solution relied on X-ray 
observations. Using $\sim$1 year of observations, consisting of over 60,000 photons, a timing solution was obtained 
based solely on gamma rays \citep{LATGEMINGA}. Geminga shows many similarities to Vela and the Crab. The phase-averaged 
spectrum is also well represented by a power law with exponential cutoff, with a hard spectral index and a cutoff energy 
between 2--3 GeV, leading to pulsed gamma rays being detected up to at least 18 GeV~\citep{LATGEMINGA}. Detailed phase-resolved 
spectroscopy shows an evolution of the spectral parameters with phase and appears to indicate that there is emission 
coming from the pulsar at all rotational phases, favoring, once again, outer-magnetospheric emission models~\citep{LATGEMINGA}.
 
The remaining EGRET pulsars, PSRs J1057$-$5226, J1709$-$4429, and J1952+3252, while still bright, were 
not as bright as Vela, the Crab, and Geminga. LAT observations of these pulsars shed light on some of the key questions 
left over from the EGRET era. All three pulsars, once again, can be fit with a power law with a
simple exponential cutoff. This contradicts earlier EGRET results that indicated that PSR\,J1709$-$4429 could be 
fit with a broken power law and PSR\,J1952+3252 showed no signs of a cutoff below 30 GeV~\citep{LAT3EGRETpulsars}. It is 
interesting to note that the conclusion about the EGRET spectrum of PSR\,J1952+3252 was based on the detection of 2 photons 
above 10 GeV.

Finally, although not detected by EGRET, PSR\,B1509$-$58 was seen by the COMPTEL instrument, and is therefore one of the 
7 gamma-ray pulsars detected by CGRO. More recently, its detection has also been reported by the {\it AGILE} 
collaboration~\citep{AgilePulsars}. Using 1 year of data, the LAT was able to detect pulsations from PSR B1509$-$58 up to 1~GeV, 
and confirmed that, unlike the EGRET-detected pulsars, PSR B1509$-$58 has an energy spectrum that breaks at a few tens 
of MeV~\citep{LATB1509}.

The high precision phase-resolved spectral measurements made possible with the LAT will be critical for theoretical modeling 
efforts, which must confront these new data. With the simple question of polar cap vs. outer magnetosphere origin now 
largely resolved, the important questions become more subtle: Where exactly in the outer magnetosphere is the 
acceleration occurring? Which magnetosphere geometry is appropriate (e.g. vacuum dipole or force-free magnetosphere)?

\section{Young Pulsars Found Using Radio Ephemerides}

\begin{figure}
\sidecaption[t]
\includegraphics[width=64mm]{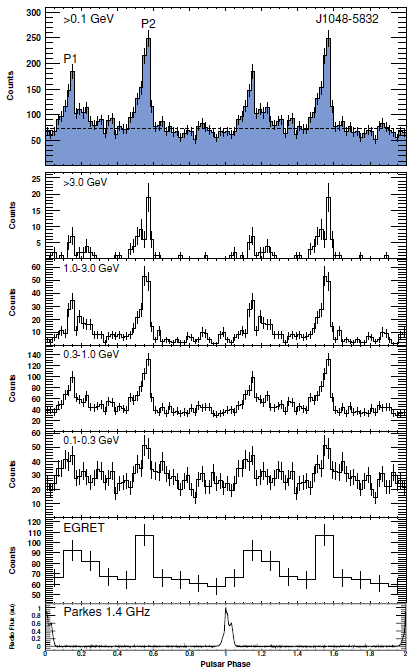}
\caption{Folded light curves of the young energetic pulsar PSR\,J1048$-$5832 \citep[from][reproduced by permission of the AAS]{LATVelaLike}. 
The second panel from the bottom shows the EGRET light curve. The {\it 
Fermi} LAT data allow us not only to confirm the marginal EGRET detection (note that the 
peaks line up), but also to study much finer time scales, as well as the energy 
evolution of the light curve.}
\label{fig:J1048}       
\end{figure}

In addition to the 7 young (or middle-aged) gamma-ray pulsars previously detected by CGRO, the 
LAT has also detected gamma-ray emission from an additional dozen or more ``young" (non-millisecond) 
radio-selected pulsars. PSR\,J2021+3651 holds the distinction of being the first new gamma-ray pulsar in 
the post-EGRET era. The pulsations were detected with the LAT during the commissioning phase of the 
instrument~\citep{PSR2021LAT}, although the original discovery of the gamma-ray pulsar was independently 
reported using {\it AGILE} data~\citep{Halpern2008}. Other pulsars detected early in the mission include 
PSR\,J1028$-$5819, shown to be at least partly responsible for the EGRET source 3EG\,J1027$-$5817, the single-peaked 
PSR\,J2229+6114 in the ``Boomerang" pulsar wind nebula (PWN)~\citep{LATVelaLike}, and the very 
energetic PSR\,J0205+6449, in SNR 3C 58~\citep{LATPSR0205}. Several of the newly-detected gamma-ray pulsars were 
already proposed as marginal EGRET detections, including PSRs J1048$-$5832 and J0659+1414. Figure~\ref{fig:J1048}, 
for example, shows the folded light curves of PSR~J1048$-$5832, including that generated with EGRET data. While 
the significance of the EGRET pulsation is clearly limited by the much lower statistics, the perfect alignment of 
the peaks with the LAT profile confirms that this was, indeed, a real detection, as originally reported by ~\citet{EGRETB1046m58}. 
Other young pulsars now seen by the LAT were originally discovered in radio searches of EGRET unidentified 
sources, and thus proposed as the energetic radio counterparts of the known gamma-ray sources (e.g. PSRs J2021+3651 
and J2229+6114). Many, however, had no previous gamma-ray associations. While the brightest new gamma-ray pulsars 
(particularly those coincident with formerly-unidentified EGRET sources) could have been detected in blind searches 
of LAT data (or searching around the extrapolation of the original radio timing solution), the detection of 
pulsations from fainter gamma-ray pulsars (e.g. PSR\,J0205+6449) requires contemporaneous 
phase-connected timing solutions spanning the entire LAT data set. In anticipation of such needs, a 
comprehensive pulsar monitoring campaign (known as the Pulsar Timing Consortium) was set up, prior to launch, 
between the LAT collaboration and the major radio telescopes, to ensure periodic monitoring of hundreds of 
pulsars with large spin-down energies, with the goal of providing the necessary ephemerides~\citep{Smith08}.

\section{Millisecond Pulsars}

At first glance, millisecond pulsars (MSPs) might not seem like great candidates for 
gamma-ray emission. After all, they are several orders of magnitude older than the 
gamma-ray bright young pulsars and their surface magnetic fields are about four orders 
of magnitude weaker.  On the other hand, their very rapid rotation rates give them open 
field line voltages that are competitive with the young pulsars and their magnetic 
fields at the light cylinder ($B_\mathrm{LC}$) are at about the median value for the young gamma-ray 
pulsars. This, plus the marginal detection of PSR J0218+4232 with EGRET \citep
{Kuiper2000}, gave some reason to be optimistic.   One particularly prescient paper 
\citep{Story07} used a detailed population study based on the pair-starved polar cap 
model to predict that the LAT should be sensitive enough to detect tens of gamma-ray 
millisecond pulsars, most of which should be radio quiet and thus form a high-latitude 
population of unidentified gamma-ray pulsars. They also pointed out that, since the high 
latitude regions have been very poorly covered by millisecond pulsar surveys so far, 
radio searches of LAT point sources with pulsar-like spectra should be an efficient way 
to find new MSPs.

Over the first 18 months of the \textit{Fermi} mission, it has become abundantly clear 
that millisecond pulsars are a significant contributor to the population of high 
latitude gamma-ray sources being detected with the LAT. Figure~\ref{fig:msps} shows the distribution, 
in Galactic coordinates, of all the MSPs detected to date with {\it Fermi} LAT. We describe these discoveries in 
the following subsections.

\subsection{Radio MSPs}

The first LAT results on MSPs came from folding the gamma-ray data using radio 
ephemerides for the $\sim 72$ field MSPs (i.e. $P<30$ ms and outside of the globular 
cluster system). Within the first 8 months of data taking, significant gamma-ray 
pulsations were discovered from 8 MSPs, including confirmation of the EGRET detection of 
PSR J0218+4232 \citep{PSR0030,MSP}. In addition to the 8 pulsed detections, it was noted 
that there were significant LAT point sources positionally coincident with 5 other MSPs 
\citep{MSP}. With continued exposure accumulating, 3 of those 5 now have reported 
pulsation detections above the 5 $\sigma$ significance level: PSR J0034$-$0534 \citep
{PSRJ0034}, PSR B1937+21 and PSR B1957+20 \citep{k10}, bringing the total number of 
radio-timed millisecond gamma-ray pulsars to 11.

The initial 8 MSP discoveries tended to resemble the normal pulsar population in most of 
their characteristics, including the peak separations, fraction that showed single vs. 
double peaks, and radio lags. This led to the conclusion that MSPs had essentially the same 
gamma-ray emission mechanism operating in the outer magnetosphere as the young pulsars, as suggested by
the similar values of $B_\mathrm{LC}$.  Interestingly, the three latest discoveries all have 
gamma-ray light curves that appear to have peaks that are \textit{aligned} in phase with the 
radio pulses. This characteristic is very rare among the normal pulsars, with the 
primary counter-example being the Crab pulsar (where the radio peaks overlap the gamma-ray ones, even though they aren't perfectly aligned, as described above). It has been suggested that these are 
cases where the gamma-ray and radio emission are coming from nearly co-located regions 
of the magnetosphere and that both result from caustic formation \citep{PSRJ0034}.
 
\subsection{Searches of LAT Unassociated Sources}
 
\begin{figure}[b]
\includegraphics[width=117mm]{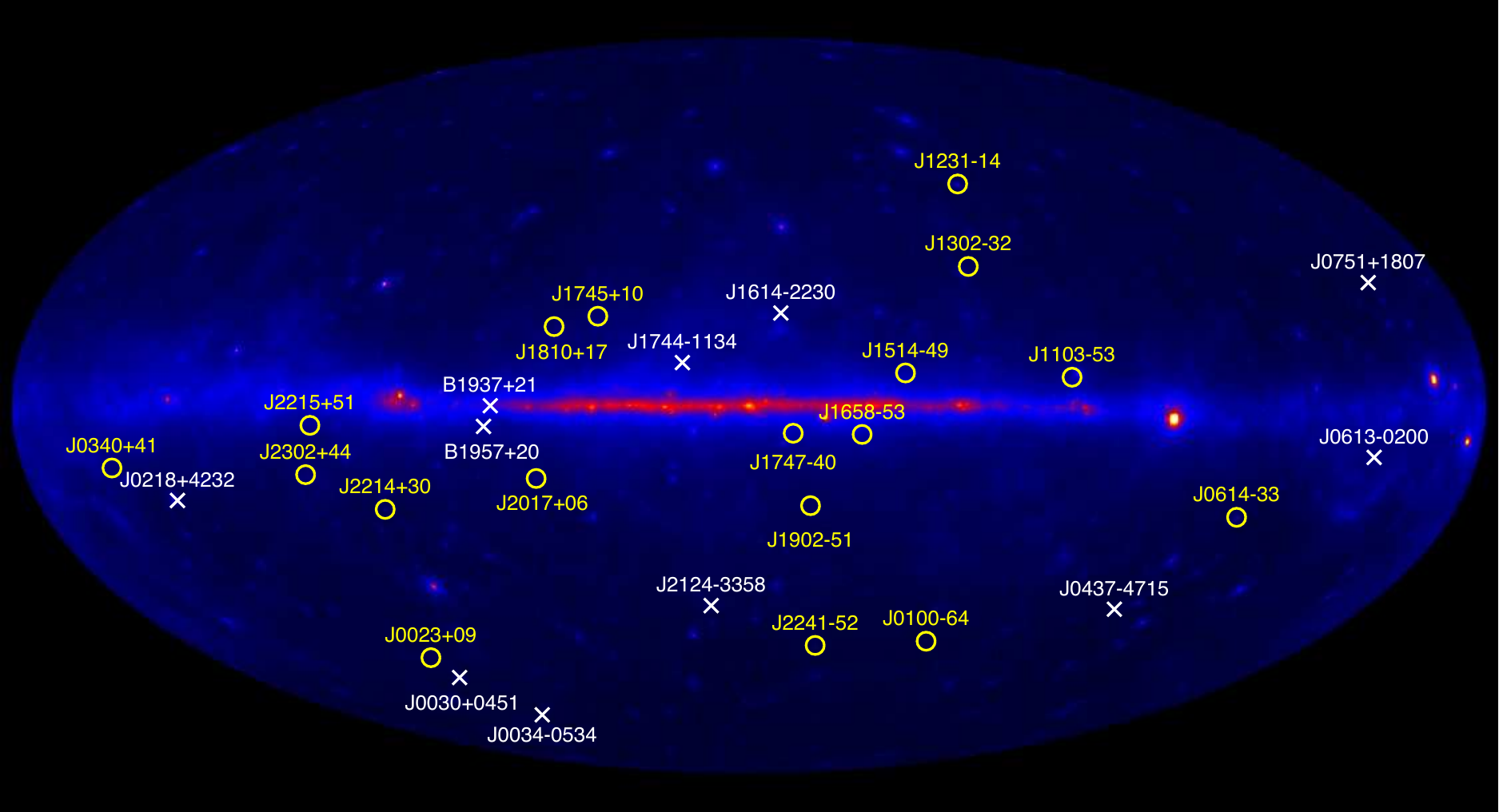}
\caption{Sky map, in Galactic coordinates, showing millisecond pulsars detected with the 
{\it Fermi} LAT. The background image is made from 16 months of LAT data (2008-08-04
through 2009-12-02) with $E>100$ MeV.
The white crosses mark the 11 previously known radio pulsars found to be 
gamma-ray pulsars with the LAT. The yellow circles indicate the 18 new radio MSPs 
discovered in searches of pulsar-like LAT unassociated sources.}
\label{fig:msps}       
\end{figure}

As mentioned earlier, a promising technique for discovering new MSPs is to perform radio 
searches in the direction of gamma-ray point sources that have pulsar-like 
characteristics (e.g. lack of variability and exponentially cutoff spectra).  This 
technique was used, with modest success, on many of the EGRET unidentified sources 
\citep[for example]{crh+06,cml05,kjk+08}.  These searches were challenging because the 
EGRET error boxes were many times larger than a typical radio telescope beam, requiring 
many pointings to cover the source region. With the LAT, the unassociated source 
localizations are a much better match to radio telescope beam sizes and can generally be 
searched in a single pointing. The \textit{Fermi} Pulsar Search Consortium (PSC) was 
conceived to organize search observations of LAT-discovered pulsars and unassociated 
sources using several large radio telescopes around the world. Thus far, over 100 LAT 
unassociated sources, mostly at high Galactic latitudes (See Figure~\ref{fig:msps}), have 
been searched at 350, 820, or 1400 MHz resulting in the discovery of 18 new millisecond pulsars \citep{ran10}. 
These searches are ongoing, and there is no apparent strong correlation 
between the gamma-ray and radio fluxes of these pulsars, so more discoveries can 
be expected as fainter LAT unassociated sources\footnote{The 1FGL catalog, compiled with 11 months of data, 
lists 630 unassociated sources and many more are expected as the LAT pushes down in sensitivity. Note, however, 
that AGN (which represent about half of the current associations) will likely comprise a significant fraction 
of these.} are searched.

These discoveries represent a $\sim$25\% increase in the number of known millisecond 
pulsars outside of the globular clusters, which is an impressive achievement considering 
the enormous effort that has gone into radio MSP searches over the last three decades. 
The new pulsars include several highly interesting sources. Five of them are so-called 
``Black Widow'' pulsars, with minimum companion masses of 0.01--0.05 $M_\odot$ and one 
other has a more typical mass companion, but exhibits radio eclipses. These more than 
double the known population of such pulsars in the field of the Galaxy and will be 
important systems for understanding the formation and evolution of millisecond pulsars 
as well as excellent systems to look for unpulsed gamma-ray emission from intra-binary 
shocks. Several others are very bright with sharp radio profiles that have the potential 
to be important additions to pulsar timing array projects that seek to detect nanoHertz 
gravitational waves via their effect on pulse arrival times \citep{haa+10}.

Since all of these pulsars are positionally coincident with LAT gamma-ray sources, it is 
expected that once sufficiently accurate timing models are available, they will all be 
found to be gamma-ray pulsars and, indeed, LAT pulsations have already been discovered 
for the first 3 of the new MSPs \citep{ran10}.
 
\subsection{Globular Cluster MSPs}

Although there are some 140 pulsars known in globular clusters\footnote{\url{http://
www.naic.edu/~pfreire/GCpsr.html}}, most of which are MSPs, there have been no reported 
gamma-ray pulsations from individual millisecond pulsars in globular clusters with the 
LAT. However, in at least 8 cases, there are point-like gamma-ray sources spatially 
coincident with globular clusters \citep{47Tuc,khc10,GCPop}. In general, these 8 LAT 
sources are consistent with being the combined emission of a population of millisecond 
pulsars in each cluster. Most have spectra that show an exponential cutoff in the few 
GeV range, as seen with MSPs, but for a couple the significance of the cutoff is too low 
for it to be considered evidence for an association of the gamma-ray source with the 
cluster.  The fluxes are largely consistent, within the substantial uncertainties, with 
estimates of the total number of MSPs in each from from radio and X-ray observations. 
However, in three cases, there are no known MSPs in clusters with associated LAT 
sources, providing a strong motivation for deeper radio pulsar searches of those 
clusters.

It is worth noting that the {\it AGILE} collaboration reported the detection of pulsations from 
PSR\,J1824$-$2452, in the globular cluster M28 \citep{AgilePulsars}, but the detection 
was marginal and appeared in only one subset of the {\it AGILE} data.  Thus far, this 
result has not been confirmed by \textit{Fermi}. In general, the detection of individual 
gamma-ray pulsars in globular clusters will likely be difficult because of the typically large distances to the 
clusters (4--12 kpc for the likely LAT-detected clusters) and because of the background 
provided by the rest of the pulsars in the cluster. However, in cases where there is 
one pulsar (like PSR J1824$-$2452) that has a very large $\dot{E}$, it may outshine the 
rest of the pulsars in the cluster and be detectable individually. Searches with the LAT are 
ongoing, using radio timing models for a large number of individual pulsars in globular clusters.

\section{Blind Periodicity Searches}

\begin{figure}[b]
\includegraphics[width=117mm]{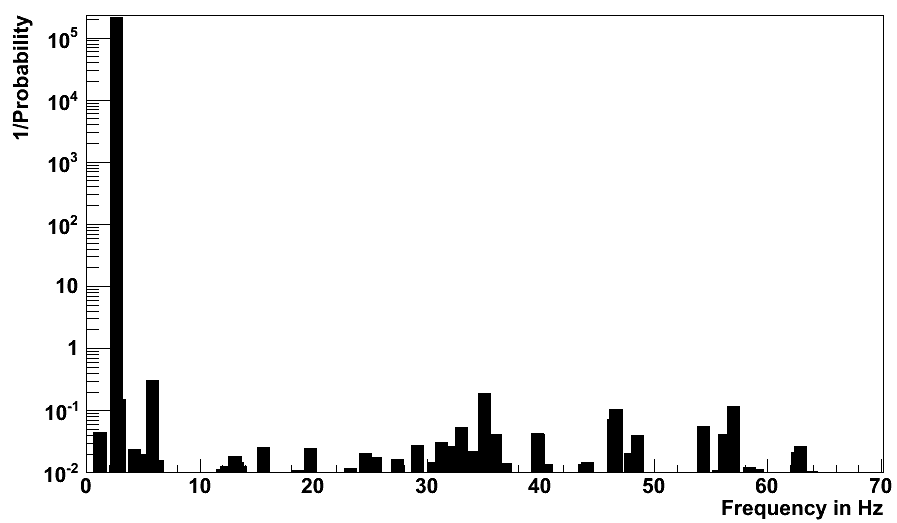}
\caption{Results from a blind search on a formerly unassociated LAT source~\citep[now PSR\,J1957+5033;][]{BSP2}, 
indicating the presence of a highly significant pulsation at 2.668 Hz. The FFT has been computed using the differences between 
binned photon arrival times up to a maximum difference of 262,144~s, and the power at each frequency has been 
normalized to represent the inverse of the probability that it could be due to a random fluctuation, as described 
in \citet{Ziegler2008}. Note that the logarithmic scale results in the majority of the 33,554,432 FFT bins not showing up 
in the figure.}
\label{fig:FFT_Example}       
\end{figure}

As described in previous sections, it was long thought that many of the EGRET unidentified sources 
could, in fact, be pulsars---in particular radio-quiet pulsars like Geminga. Previous attempts to carry out 
blind searches on EGRET data using coherent FFT techniques were unsuccessful 
\citep[e.g.][]{Chandler2001}. The sparse data sets and sensitivity to timing irregularities make
such searches incredibly challenging. A new technique was developed to try and ameliorate the 
problem, by calculating the FFT of the time differences (instead of times of arrival) of events. 
Time differences are calculated between all events in the time series with respect to events lying 
within a relatively short sliding window ($\sim$weeks). The lower frequency resolution of the resulting FFTs 
make these searches less sensitivite to frequency shifts (such as those caused by the spindown of 
the pulsar), while at the same time resulting in great savings in computational time~\citep{Atwood2006}.
This new time-differencing technique was shown to work with EGRET data~\citep{Ziegler2008}, and has since 
proven extremely successful with the LAT data, leading to the discovery, so far, of 24 pulsars found in 
blind searches~\citep{BSP,BSP2}. Figure~\ref{fig:FFT_Example} shows an example of the output from a 
successful blind search of a formerly unassociated LAT source, now identified as PSR\,J1957+5033~\citep{BSP2}. 
After determining that the highly significant peak at 2.668 Hz is promising, standard pulsar packages such as 
PRESTO\footnote{{\tt http://www.cv.nrao.edu/$\sim$sransom/presto/}}~\citep{RansomThesis}, 
and tempo2~\citep{hobbs06}, are used to refine the result and obtain a final timing solution for the pulsar.

Most of the initial 16 pulsars found in blind searches of LAT data were associated with 
formerly unidentified EGRET sources. In fact, only 3 of the 16 had no EGRET counterpart~\citep{BSP}.
Many of these sources were long-suspected of hosting pulsars, including 3EG\,J1835+5918, the brightest unidentified 
EGRET source off the Galactic plane, which was even dubbed the `Next Geminga' \citep{Halpern2007}. Six out of the 
sixteen pulsars were discovered by assuming a counterpart position derived from observations at other wavelengths (mostly X-ray), instead of the less precise 
LAT position. A prime example is the discovery of PSR\,J1836+5925 powering 3EG\,J1835+5918~\citep{LATPSR1836}.
More recently, the last 8 pulsars found in blind searches have mostly been found from 
newly-discovered LAT sources, with no corresponding EGRET counterpart, except in some cases where 
the EGRET source might have been confused and is now being resolved into multiple separate gamma-ray 
sources by the LAT~\citep{BSP2}.
 
Although radio beaming fractions for MSPs appear to be large \citep{kxl+08}, 
there are still expected to be radio quiet 
millisecond pulsars detected as point sources with the LAT. A discovery of a radio-quiet 
MSP in a blind search would be an important result.  Unfortunately, the parameter space 
that needs to be searched is vast.  For the case of binary MSPs the problem may be 
essentially intractable.  However, about 25\% of MSPs are isolated, including at least 
two of the LAT-detected radio MSP.  For these pulsars the search is daunting, but not 
impossible. On the plus side, MSPs have low period derivatives and are extremely stable 
rotators, so the pulse will remain phase coherent for a long integration time.  
Counteracting this is the fact that the fast spin rates require that the pulsar position 
be known very precisely ($\sim 0.1\  \mathrm{arcsec}$).  For a typical LAT point source 
position uncertainty of 3 arcmin, this requires $3.2\times10^{6}$ trial positions to 
cover the region, and each trial position requires a search over frequency and frequency 
derivative.

Current efforts on blind searches of LAT data are concentrating on both searching deeper for young and middle 
aged pulsars as well as expanding the search parameter space to include isolated MSPs.

\section{Pulsar Timing with the LAT}

Pulsar timing is a powerful technique that involves fitting a model to measured pulse arrival times 
that can account for every rotation of the neutron star over a time span of years \citep
[chap. 8]{Handbook}.  Of course, such timing yields extremely precise measurements of 
the spin period and spindown rate of the neutron star, quantities from which estimates 
of the magnetic field, age, and energy loss rate of the pulsar can be derived. In 
addition, because of the motion of the Earth around the solar system barycenter, the 
pulse arrival times are highly sensitive to the pulsar position on the sky.  Once those 
major effects are accounted for, timing is sensitive to a host of other parameters of 
the system including binary orbital parameters, timing noise, glitches, and even proper 
motion and parallax in some cases.

Traditionally, pulsars have been discovered and timed using radio telescopes. 
Working in the gamma-ray band, EGRET was 
not very effective for pulsar timing both because of its limited sensitivity and because 
of its pointed viewing plan that meant that most pulsars were only observed for a few 
2-week observations scattered over the mission. The situation is completely different with 
\textit{Fermi}, which now has both the sensitivity to detect a large number of pulsars 
and a sky survey viewing plan that allows observations of every pulsar in the sky continuously.  For 
most of the 24 blind search pulsars, timing using the LAT data is the only option since 
they are undetectable or extremely faint at radio wavelengths. In addition there are some very faint radio pulsars, such as PSR J1124$-$5916 where the observation time required to do radio timing is prohibitive, but which can be readily timed with the LAT.

There are several key differences between pulsar timing with the LAT and radio pulsar 
timing. First, the satellite is not affixed to the Earth, like a ground-based radio 
telescope.  Second, the data are very sparse, with often fewer than 100 photons being used 
to make a pulse time-of-arrival (TOA) measurement. The first issue is dealt 
with by transforming the photon arrival times as observed at the satellite to a 
fictional observatory at the geocenter, thus removing the effects of the spacecraft 
motion on the measurement. The second difference drives one to adopt a TOA measurement 
technique different than the traditional radio method of cross correlating a folded 
pulse profile with a high signal to noise binned template. Instead, TOAs are determined 
by a maximum likelihood fit to the offset between the measured photon times and an 
analytic template profile \citep{BlindTiming}.

\begin{figure}
\sidecaption
\includegraphics[width=75mm]{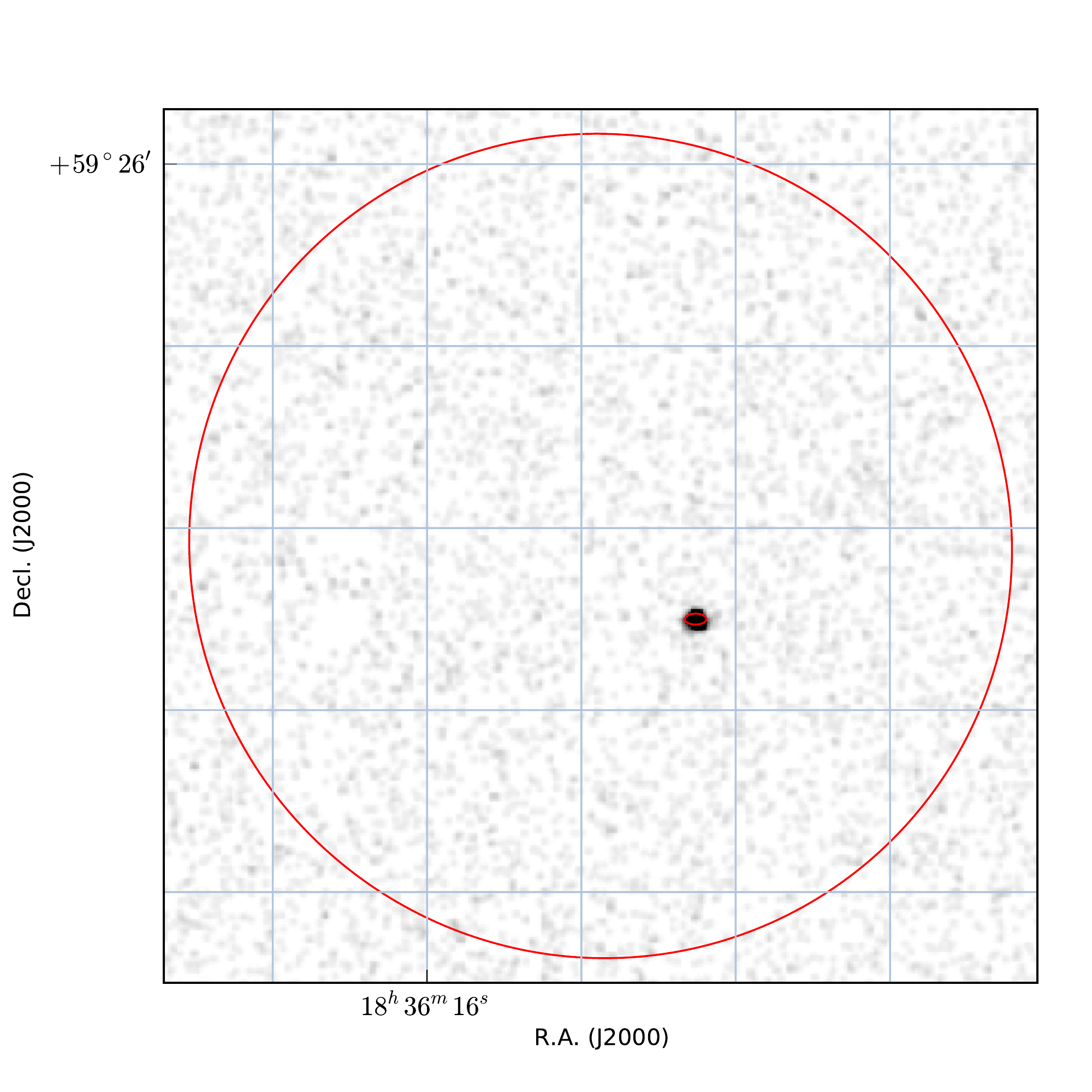}
\caption{Comparison of position determinations of PSR J1836+5925. The large ellipse
(0.45 arcmin semimajor axis) is the 95\% confidence region from positional analysis 
of 18 months of LAT data (M. Kerr, private communication). The small ellipse (0.8 
$\times$ 0.4 arcsec) is from the pulsar timing model fit over the same interval 
\citep{BlindTiming}.  The background image is a Chandra X-ray image showing the 
point source at the location of the pulsar.}
\label{fig:pos}       
\end{figure}

What is impressive is that even with so few photons, timing models can be determined for 
most detectable LAT pulsars with RMS residuals of order a millisecond using TOAs spaced 
by a few weeks.  This enables arcsecond position determinations as shown in Figures~
\ref{fig:pos} and \ref{fig:J1023} (right panel).

In addition to these precise positions that enable multiwavelength counterpart 
identifications, pulsar timing with the LAT has provided spindown measurements for the 
gamma-ray selected pulsars, detection and measurement of glitches, and studies of the 
timing noise observed in these systems. The precise long-term timing models are also 
critical for other studies such as blanking a pulsar to remove confusion in the study of 
a nearby source, as was required for Cygnus X-3 \citep{CygX3} or
searches for off-pulse emission, such as from an SNR or PWN~\citep{gl10}.

\section{Multiwavelength Connections}

The 24 blind-search pulsars were all discovered in gamma-ray searches and thus are 
gamma-ray selected pulsars, but targeted radio observations are required to determine if 
they are also radio quiet, or could have been discovered in radio surveys independently.  
Radio detections also yield distance estimates from dispersion measure, information on 
the emission region from radio to gamma-ray offset, and geometry from radio polarization 
studies. In addition, the population statistics of radio quiet vs. radio loud gamma-ray 
pulsars have important implications for gamma-ray emission models.

\begin{figure}[t]
\sidecaption[t]
\includegraphics[width=75mm]{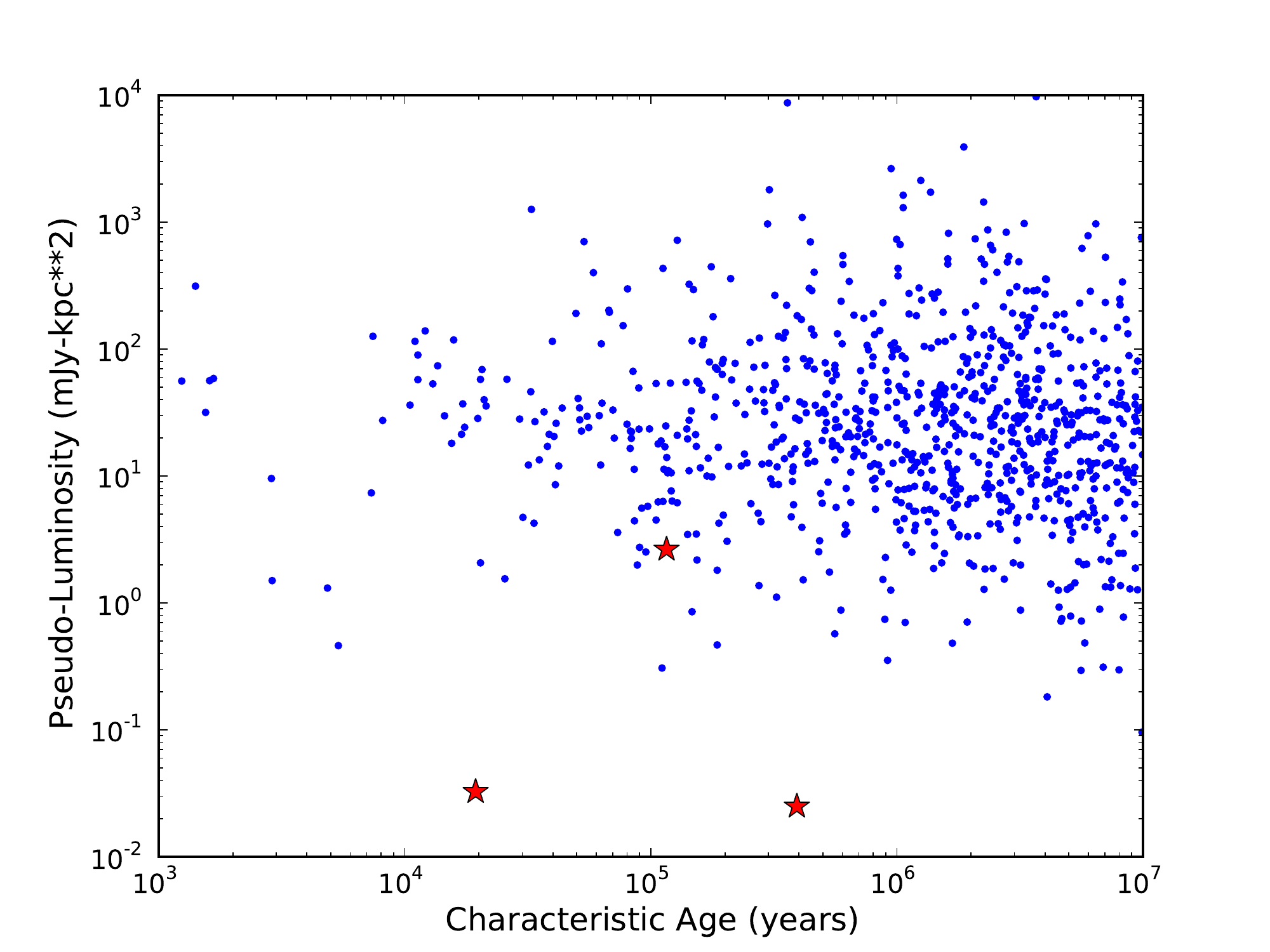}
\caption{Pseudo-luminosities of the gamma-ray selected pulsars that have since been 
detected as radio pulsars (red stars), compared to the general population of radio pulsars (blue dots).}
\label{fig:lumfig}       
\end{figure}

Deep radio searches have now resulted in the detection of radio pulsations from three of the 24 
blind search pulsars, with strong upper limits on the others \citep{BlindTiming,BSP2}.  The pulsations from 
J1741$-$2054 were found in archival Parkes Multibeam survey data and confirmed using the 
Green Bank Telescope (GBT) \citep{Camilo2009}. For J2032+4127, the pulsations were discovered using the GBT 
\citep{Camilo2009}. The third radio pulsation discovered was from PSR J1907+0602 \citep
{MGRO} using a very deep observation with the 305-m Arecibo telescope. The detections 
provide distance estimates from the dispersion measure, which allow conversion of the 
radio fluxes into pseudo-luminosities. As shown in Figure \ref{fig:lumfig}, two of these pulsars are exceptionally faint, with luminosities about an order of magnitude lower than the faintest radio-discovered young pulsars. This is forcing a reevaluation of what is meant by a `radio quiet' pulsar.

\begin{figure}
\includegraphics[width=117mm]{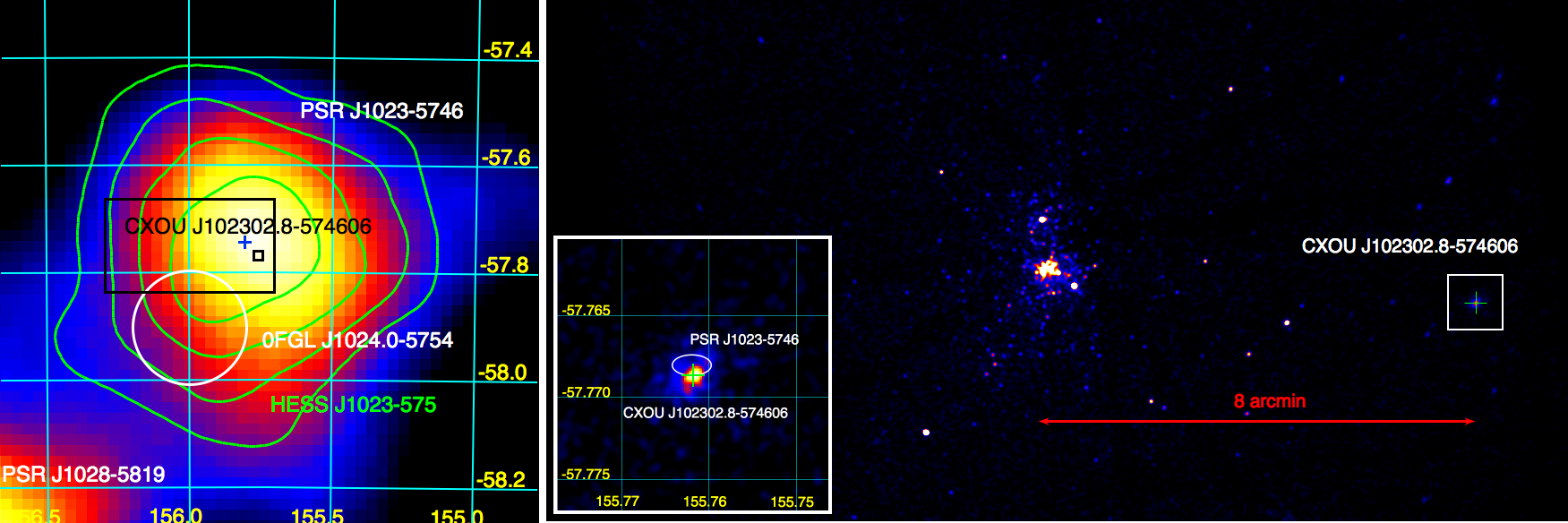}
\caption{{\bf Left --} {\it Fermi} LAT counts map of the region around PSR\,J1023$-$5746. 
The green contours represent the HESS significance. {\bf Right --} {\it Chandra} X-ray 
image of the Westerlund 2 cluster. The X-ray counterpart of PSR\,J1023$-$5746 is 
approximately 8 arcminutes away from the core of the cluster. Inset: Zoomed-in image of a 1 
square arcminute region around the location of the pulsar. Note that the 95\% (statistical) error 
ellipse obtained from pulsar timing (shown in white) overlaps with the X-ray source. Figures 
from \citet{BSP2}, reproduced by permission of the AAS.}
\label{fig:J1023} 
\end{figure}

Observations of PWNe at TeV energies go back to the very first firm detection of emission from the Crab nebula~\citep{Weekes89}. 
Since then, over 100 TeV sources have been detected\footnote{For an up-to-date catalog of TeV sources, 
see {\tt http://tevcat.uchicago.edu/}} and more than half of these have associated LAT sources~\citep{1FGL}, which is 
perhaps not altogether surprising given that the energy ranges of the LAT and ground-based Cerenkov detectors overlap.
PWNe represent the largest class of Galactic TeV sources. In fact, the first unidentified TeV
source, discovered by the HEGRA telescope in the Cygnus OB2 region, is associated with PSR J2032+4127, one of the pulsars 
found in blind searches of LAT data~\citep{BSP,Camilo2009}. HESS observations of the Galactic plane uncovered a large number of 
unidentified TeV sources, and many of these are thought to be associated with PWNe. In some cases, the discovery of new LAT 
pulsars coincident with known TeV sources can put into question previous interpretations of the TeV emission. Figure~\ref{fig:J1023}, 
for example, shows the positional coincidence of the highly energetic pulsar PSR\,J1023$-$5746~\citep{BSP2} with the bright TeV source, 
HESS\,J1023-575. Located in the vicinity of the young stellar cluster Westerlund 2, the TeV emission from this source was previously 
thought to be due mainly to the wind interaction from massive stars~\citep{HESSJ1023}. The presence of such a pulsar, however, must 
lead to a re-examination of such conclusions. Furthermore, the identification of the counterpart (right panel in Figure~\ref{fig:J1023}) 
shows that the association with the Westerlund 2 cluster is highly questionable.

At higher energies still, the Milagro observatory detected significant ($> 5\sigma$) TeV emission at a median energy of 35 TeV
from the location of 6 gamma-ray pulsars detected by the LAT, and evidence for emission ($3-5 \sigma$) from the location of 
an additional 8 sources from the Bright Source List~\citep{Milagro2009}. Four of those sources are gamma-ray pulsars, and 
two more are associated with supernova remnants.

This strong connection between young energetic GeV pulsars and their TeV PWNe can play an important role not only in 
understanding the nature of the emission from such sources, but also as a means to identify likely candidates for gamma-ray
pulsars, ultimately leading to the identification of both TeV and GeV sources.

X-ray observations of gamma-ray pulsars and pulsar candidates are particularly important. First, the precise positions of 
neutron star candidates allow for more sensitive blind searches to take place (as in the case of PSR\,J0007+7303 or PSR\,J1836+5925). 
Secondly, for those pulsar candidates found using the less precise LAT position, X-ray positions can serve to refine the 
candidate and determine whether it is a real detection. In 4 out of the original 16 pulsars discovered in blind searches, a short 
observation with the Swift satellite was enough to identify a plausible X-ray counterpart which resulted in a much higher 
significance of the pulsation~\citep{BSP}. In several other cases (e.g. Gamma Cygni SNR, Cygnus OB2 association), archival 
observations could be analysed in search of the best possible counterpart.

\begin{figure}[h]
\begin{center}
\includegraphics[width=100mm]{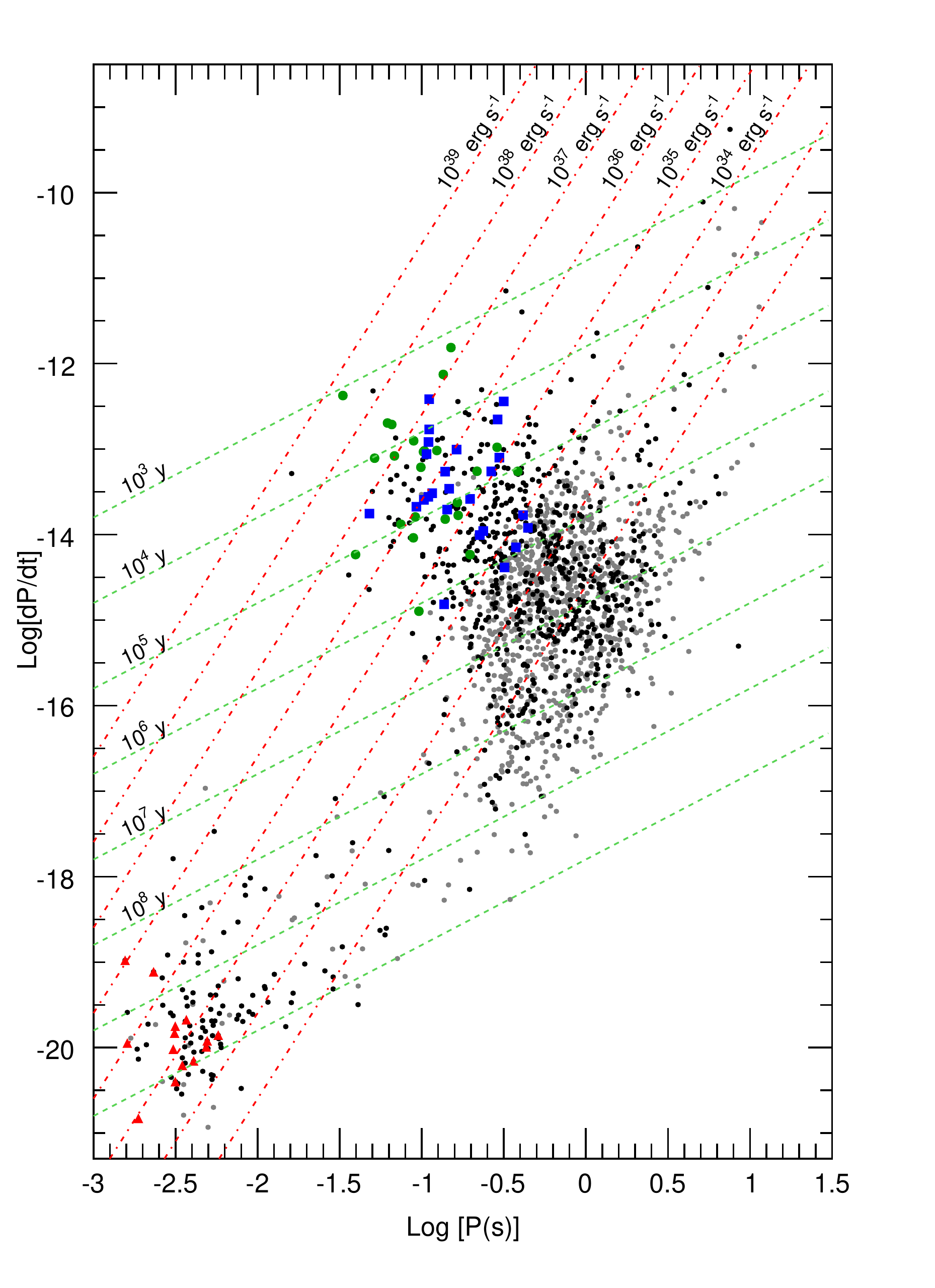}
\end{center}
\caption{Period-Period Derivative diagram showing the LAT-detected pulsars. Included are 
24 young or middle-aged radio-timed pulsars (green circles), 25 gamma-ray selected 
pulsars (blue squares), where all but Geminga were discovered in LAT blind searches, and 
14 millisecond pulsars (red triangles), for a total of 63 gamma-ray pulsars. Note that this does not include 15 of the radio millisecond pulsars discovered in searches of LAT unassociated sources, essentially all of which can be expected to be detected as gamma-ray pulsars once their timing models are well determined.}
\label{fig:ppdot}       
\end{figure}

\section{The LAT Pulsar Population}

A Period-Period Derivative diagram showing all 63 gamma-ray pulsar detections made with 
the LAT to date is shown in Figure \ref{fig:ppdot}. This is an update of Figure 2 from 
the {\it Fermi} LAT First Pulsar Catalog~\citep{PulsarCatalog}, which summarizes the 
characteristics of the 46 gamma-ray pulsars detected with the LAT in the first 6 months 
of the {\it Fermi} mission. The LAT-detected pulsars generally have high values for the 
detectability metric $\dot{E}^{1/2}/D^2$ and large $\dot{E}$ and $B_\mathrm{LC}$. Which 
one of these is really telling us about the gamma-ray emission physics at work in these 
sources remains to be seen.  With a large number of detections spanning a range of 
$\dot{E}$ from $10^{33.5}$ to $>10^{38}$ erg s$^{-1}$, we can start to address the 
evolution of gamma-ray luminosity (i.e. efficiency) with $\dot{E}$. Unfortunately, the 
large distance uncertainties for most pulsars combined with the model-dependent 
uncertainty in the beaming factor (see below) prevent strong conclusions from being drawn at present \citep{PulsarCatalog}.

The spectra of LAT pulsars are well characterized by exponentially cutoff power laws 
with photon indices near 1.5. The cutoff energies are in the 1--4 GeV range with a 
small number of outliers on the high and low side. The observed pulse profiles 
frequently evolve with energy, but generally fall into one of three categories: two 
peaks separated by $\sim 0.4-0.5$ in phase, two overlapping peaks separated by $\sim 
0.2$ in phase, and single peaked profiles.  Most of the LAT pulsars are consistent with 
being 100\% pulsed in the gamma-ray band.  However, a few (e.g. Geminga and PSR 
J1836+5925) seem to show magnetospheric emission across all rotational phases.  In other 
cases, an analysis of the `off-pulse' region of pulse phase reveals GeV emission from a 
pulsar wind nebula, typically with a much harder spectrum than that of the pulsar 
itself. A review of LAT observations of PWNe is presented elsewhere in this volume 
\citep{gl10}.

The large number of radio and gamma-ray selected pulsars found with the LAT, combined 
with deep radio searches of the new gamma-ray selected population will enable population 
studies that will help shed light on the beaming fractions in the two bands and test 
predictions of the various models for the emission region geometries. One such early 
study \citep{rmh10} finds that the radio beaming fraction is near unity for the the 
highest $\dot{E}$ pulsars and decreases to $\sim 0.5$ for the lower $\dot{E}$ gamma-ray 
pulsars, implying that very high-$\dot{E}$ pulsars may produce their radio emission in 
the outer magnetosphere. If confirmed, this would have major implications.


The current challenge is to use the abundance of well-measured light curves to constrain the geometry of the 
emitting region and the relevant magnetospheric physics.  The favored approach is to choose an emission region 
location (e.g. polar cap (PC), outer gap (OG), or two-pole caustic (TPC)), combine it with an assumed magnetic 
field geometry and compute an `atlas' of predicted gamma-ray light curves that can be compared with observations. 
This has been done for vacuum dipole field geometries \citep{Watters09}, as well as for numerically-modeled `force-free' 
geometries \citep{bs10b}. Other groups have specifically targeted millisecond pulsars \citep{vhg09}.
The predicted light curve morphologies are sensitively dependent on both the misalignment between the spin and 
magnetic axes of the neutron star ($\alpha$) and the viewing angle ($\zeta$) between the spin axis and the line 
of sight. Without \textit{a priori} knowledge of these angles, it can be hard to discriminate among models based 
on light curve fits. However, if the angles can be constrained by other methods, such as radio polarization measurements 
or X-ray PWN geometry, the degeneracies can be broken. An important output of these model fits is the `flux correction 
factor' $f_\Omega$, defined such that the true gamma-ray luminosity, $L_\gamma$ of a pulsar is
\begin{equation}
L_\gamma = 4 \pi f_\Omega F_\mathrm{obs} D^2,
\end{equation}
where $F_\mathrm{obs}$ is the observed gamma-ray flux and $D$ is the distance. In the EGRET era it was commonly assumed that $f_\Omega \sim 1/4\pi$, but current models predict values much closer to 1. This parameter is crucial for understanding the energetics of these systems and the efficiency ($\eta$) with which they convert rotational energy into gamma-rays. 
Recent model comparisons with a few LAT pulsar light curves \citep{rw10}
suggest that OG models with alternate field geometries are preferred
in these cases. However, other objects may be consistent with lower altitude
emission, and additional comparisons are needed to see if the data are
consistent with emission beyond the light cylinder, as suggested by the 
force-free models. With many more high quality light curves being collected by
the LAT, it should be possible to make powerful tests of these models, especially
if when angle constraints from radio and X-ray observations are available.

\section{Future Expectations}

The next few years promise a continued stream of exciting pulsar results from the LAT. 
With the very reasonable assumption that the 18 new millisecond pulsars found in radio 
searches of LAT unassociated sources will all turn out to be gamma-ray pulsars, there 
will soon be more than 75 solid gamma-ray pulsar detections. This number is not totally 
unexpected, according to several pre-launch predictions.  What is more surprising is 
that the population is divided into three essentially equal groups: young or middle-aged 
radio-selected pulsars, young or middle-aged gamma-ray selected pulsars, and millisecond 
pulsars.

Modeling the spectra, light curves, and population statistics of the LAT pulsars will
be extremely important over the next few years to turn the powerful observations into 
improved understanding of the physical mechanism for pulsar gamma-ray emission. But, 
since this is a primarily observational review, we close with a few of the important 
observational questions that we expect to be addressed in the coming years.
\begin{itemize}
\item{Are there radio quiet millisecond pulsars? This is both a great challenge for the 
observers and has very important implications for the emission mechanisms and geometry.}
\item{If the `gamma-ray binaries' LS I +61 303 and LS 5039 (see review in this volume 
\citet{hdt10}) are powered by energetic pulsars, can we detect the gamma-ray 
pulsations with the LAT?}
\item{What are the non-detections of known pulsars telling us? While the new pulsar 
discoveries have grabbed most of the attention, it may be that one or more key 
non-detections will tell us something important about what drives gamma-ray pulsars. 
However, these studies are critically reliant on accurate distance determinations, so
this is really a reminder that improved VLBA or timing parallax measurements for as 
many pulsars as possible will be of great value in increasing the science return from
 LAT pulsar studies.}
\end{itemize}

%
\begin{acknowledgement}

The \textit{Fermi} LAT Collaboration acknowledges generous ongoing support
from a number of agencies and institutes that have supported both the
development and the operation of the LAT as well as scientific data analysis.
These include the National Aeronautics and Space Administration and the
Department of Energy in the United States, the Commissariat \`a l'Energie Atomique
and the Centre National de la Recherche Scientifique / Institut National de Physique
Nucl\'eaire et de Physique des Particules in France, the Agenzia Spaziale Italiana
and the Istituto Nazionale di Fisica Nucleare in Italy, the Ministry of Education,
Culture, Sports, Science and Technology (MEXT), High Energy Accelerator Research
Organization (KEK) and Japan Aerospace Exploration Agency (JAXA) in Japan, and
the K.~A.~Wallenberg Foundation, the Swedish Research Council and the
Swedish National Space Board in Sweden.

Additional support for science analysis during the operations phase is gratefully
acknowledged from the Istituto Nazionale di Astrofisica in Italy and the Centre National 
d'\'Etudes Spatiales in France.

Pablo Saz Parkinson acknowledges support from the American Astronomical Society and the 
National Science 
Foundation in the form of an International Travel Grant, which enabled him to attend 
this conference.

Basic research in astronomy at the Naval Research Laboratory is supported by 6.1 base 
funding.

\end{acknowledgement}
%
\input{latpulsarrefs}
\clearpage{}
\end{document}

%% file: latpulsarrefs.tex
%
%
\bibliographystyle{apj}
\newcommand{\etal}{et al.}
\newcommand{\apj}{ApJ}
\newcommand{\apjl}{ApJ}
\newcommand{\apjs}{ApJS}
\newcommand{\mnras}{MNRAS}
\newcommand{\aap}{A\&A}
\bibliography{latpsr}